\documentclass{jaa}
\usepackage{natbib}
\usepackage{graphicx}

\usepackage[colorlinks=true,citecolor=blue]{hyperref}
\begin{document}\sloppy
%
\title{The upcoming 4m ILMT facility and data reduction pipeline testing}

\author{Brajesh Kumar\textsuperscript{1,*}, Vibhore Negi\textsuperscript{1, 2}, Bhavya Ailawadhi\textsuperscript{1, 2}, Sapna Mishra\textsuperscript{3}, Bikram Pradhan\textsuperscript{4}, Kuntal Misra\textsuperscript{1}, Paul Hickson\textsuperscript{5} and Jean Surdej\textsuperscript{6, 7}}
\affilOne{\textsuperscript{1}Aryabhatta Research Institute of Observational Sciences, Manora Peak, Nainital - 263001, India.\\}
\affilTwo{\textsuperscript{2}Deen Dayal Upadhyaya Gorakhpur University, Gorakhpur 273009, India.\\}
\affilThree{\textsuperscript{3}Inter University Centre for Astronomy and Astrophysics, Post Bag 4, Ganeshkhind, Pune 411007, India.\\}
\affilFour{\textsuperscript{4}Indian Space Research Organization, Antariksh Bhavan, New BEL Road, Bengaluru, Karnataka 560231.\\}
\affilFive{\textsuperscript{5}Department of Physics and Astronomy, University of British Columbia, 6224 Agricultural Road, Vancouver, BC V6T 1Z1, Canada.\\}
\affilSix{\textsuperscript{6}Institute of Astrophysics and Geophysics, University of Liège, All\'ee du 6 Ao\^ut 19c, 4000 Li\`ege, Belgium.\\}
\affilSeven{\textsuperscript{7}Astronomical Observatory Institute, Faculty of Physics, Adam Mickiewicz University, ul. Sloneczna 36, 60-286 Poznan, Poland.\\}


\twocolumn[{

\maketitle

\corres{brajesh@aries.res.in, brajesharies@gmail.com}
\msinfo{ 2021}{ 2021}

\begin{abstract}
The 4m International Liquid Mirror Telescope (ILMT) installation activities have recently been completed at the Devasthal observatory (Uttarakhand, India). The ILMT will perform continuous observation of a narrow strip of the sky ($\sim$27$'$) passing over the zenith in the SDSS $g'$, $r'$ and $i'$ bands. In combination with a highly efficient 4k $\times$ 4k CCD camera and an optical corrector, the images will be secured at the prime focus of the telescope using the Time Delayed Integration technique. The ILMT will reach $\sim$22.5 mag ($g'$-band) in a single scan and this limiting magnitude can be further improved by co-adding the nightly images. The uniqueness of the one-day cadence and deeper imaging with the ILMT will make it possible to discover and study various galactic and extra-galactic sources, especially variable ones. Here, we present the latest updates of the ILMT facility and discuss the preparation for first light, which is expected during early 2022. We also briefly explain different steps involved in the ILMT data reduction pipeline.
\end{abstract}

}]

\doinum{--}
\artcitid{\#\#\#\#}
\volnum{000}
\year{0000}
\pgrange{1--}
\setcounter{page}{1}
\lp{8}

\section{Introduction}

The initial concept of Liquid Mirror (LM) was proposed by an Italian astronomer, Ernesto Capocci in 1850 \citep{Mailly-1872}. Subsequently, in 1872, Henry Skey built the first working Liquid Mirror Telescope (LMT) with a size of 0.35m \citep[see][]{Gibson-1991-LM}. He also demonstrated that while varying the angular velocity ($\omega$) of the liquid, the focal length ($f$) of an LMT could be modified using the relation: $f=g/2{\omega}^2$ where $g$ is the local gravitational acceleration. During the early nineteenth century, Robert Wood developed LMTs of different sizes and could observe several stellar sources passing over the zenith \citep{Wood-1909}. 

The present era of LMTs began with significant contributions from Canadian teams led by Ermanno Borra and Paul Hickson. In the light of technological advances, various practical limitations were resolved. For example, implementation of air bearings for an almost friction-less rotation of the  mercury container, synchronous motor driven by an oscillator-stabilized AC power supply, use of a mylar film cover over the mirror surface, etc. \citep{Borra-1982,Borra-1992,Borra-1995}. In this way, the image degrading wavelets were substantially eliminated in order to achieve diffraction limited images \citep{Borra-1989,Hickson-2007PASP}. Later, several scientific programs were successfully conducted using LMTs as proper astronomical tools \citep{Hickson-1994ApJ, Hickson-1994, Gibson-1991}. During the past years, the Large Zenith Telescope (LZT) was the largest (6.0m diameter) fully functional LMT near Vancouver, Canada \citep{Hickson-2007LZT}. Astronomical observations were restricted with the LZT due to  inadequate site characteristics but eventually, it was used as a LIDAR facility \citep{Pfrommer-2009,Pfrommer-2012,Pfrommer-2014}.

\begin{figure*}
\centering
\includegraphics[scale=0.69]{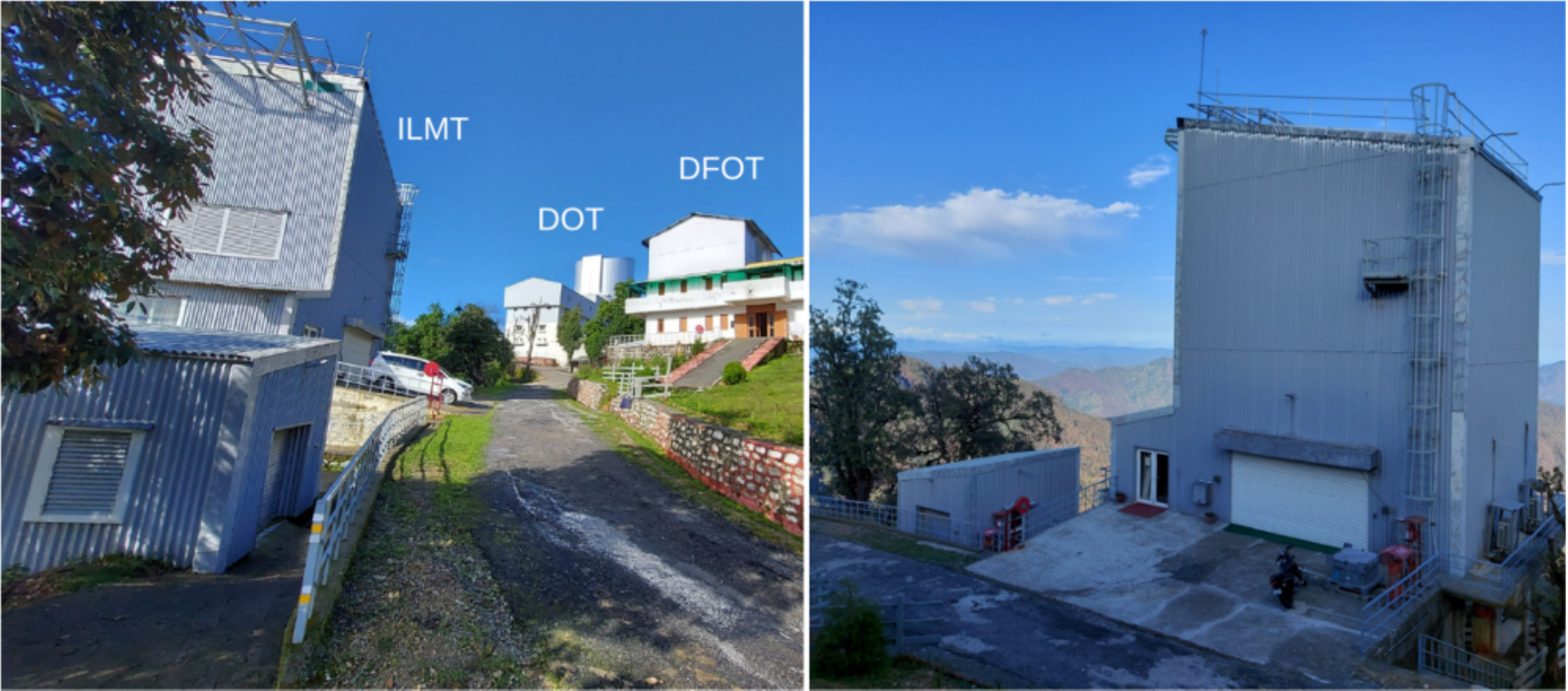}
\caption{Observing facilities at Devasthal observatory. Left panel: Locations of the 4m International Liquid Mirror Telescope (left), 3.6m Devasthal Optical Telescope (middle) and 1.3m Devasthal Fast Optical Telescope (right) are indicated. Right panel: Front view of the ILMT building with the compressor room (left), data acquisition room (middle) and main enclosure (right), respectively.}
\label{site}
\end{figure*}

\section[The ILMT project]{The ILMT project\footnote{More details about the project can be found at \url{http://www.ilmt.ulg.ac.be}} and present status}

The 4m International Liquid Mirror Telescope (ILMT) is an upcoming observing facility at the newly developed Devasthal observatory (79$^{\circ}$ 41$'$ 04$''$ E, +29$^{\circ}$ 21$'$ 40$''$ and altitude 2450m) near the Nainital city of the northern Indian state Uttarakhand. This observatory is operated by the Aryabhatta Research Institute of Observational Sciences (ARIES) under the Department of Science \& Technology, Government of India. The observatory already hosts two modern optical telescopes, considering the favourable astronomical characteristics of the site. The 1.3m Devasthal Fast Optical Telescope (DFOT) and the 3.6m Devasthal Optical Telescope (DOT) are operational since 2010 and 2016, respectively \citep{Sagar-2012,Sagar-2019,Brij-2018}. The locations of the three observing facilities are shown in Fig.~\ref{site}.

The ILMT is a collaborative project between various institutions/universities involving five countries: Belgium, India, Canada, Poland and Uzbekistan. It includes the Institute of Astrophysics and Geophysics (ULg, Li\`ege University, Belgium), the Royal Observatory of Belgium, Aryabhatta Research Institute of Observational Sciences (Nainital, India), Laval University (Qu\'ebec), University of Montr\'eal (Montr\'eal), University of Toronto and York University (Toronto), University of British Columbia (UBC, Vancouver), University of Victoria (Victoria), Pozna\'n Observatory (Poland), and Ulugh Beg Astronomical Institute of the Uzbek Academy of Sciences and National University of Uzbekistan. 
A significant fabrication of the telescope activities were accomplished by the Advanced Mechanical and Optical Systems (AMOS) company in Belgium. Before transporting the telescope to India in 2011, various technical difficulties were resolved through critical experiments carried out by UBC and ULg astronomers. It included the stiffness enhancement and spin casting of the bowl and also the verification of the mercury surface quality.

The ILMT consists of three major components, namely (i) a container/bowl to sustain the liquid, (ii) an air bearing on which the container sits, and (iii) a drive system for precise rotation. The ILMT container size is approximately 4m in diameter, in which the rotating metallic liquid mercury will act as a reflecting mirror. The telescope field of view (FOV) is $\sim$\,$27' \times 27'$, and it will perform Time Delay Integration (TDI) mode imaging (see Section~\ref{data-test}) in the SDSS $g'$, $r'$, and $i'$ spectral bands with the help of a 4k $\times$ 4k CCD camera manufactured by Spectral Instruments. To correct the aberration raised due to TDI imaging, a five lenses optical corrector will be used. A list of important parameters of the ILMT is provided in Table~\ref{tab_ILMT}.

The total integration time of the ILMT is $\sim$102 sec (single scan) and $\sim$47 deg$^2$ sky area can be monitored each night. Further, our calculations indicate that it may reach approximately 22.8, 22.3, and 21.4 mag in the $g'$, $r'$, and $i'$ bands, respectively \citep[for details, see][]{Kumar-2018MN}. It is worth noticing that since the same sky region (except for a 4 min shift in right ascension) will pass over the telescope every consecutive night, night images can be co-added to improve the limiting magnitudes.

\begin{table}
\centering
\caption{Important parameters of the ILMT and of its components \citep[see][]{Finet-2013Thesis,Kumar-2018MN}.}\label{tab_ILMT}
\begin{tabular}{ll}
\hline
Parameter                    &  Value                     \\ \hline
Mirror diameter              &  4.1 m                     \\
Focal length                 &  8.0 m                     \\
Field of view                &  $27' \times 27'$          \\
Accessible sky area          &  $\sim$\,47 deg$^2$        \\
Rotation period              &  8.02 sec                  \\
CCD camera                   &  4096 $\times$ 4096 pixels \\
CCD pixel size               &  $0.33''$ pixel$^{-1}$     \\
CCD readout noise            &  5.0 e$^{-1}$              \\
CCD gain                     &  4.0 e$^{-1}$/ADU          \\
Filters                      &  SDSS $g'$, $r'$, $i'$     \\
TDI integration time         &  102 sec                   \\
Optical corrector lenses     &  5                         \\
\hline
\end{tabular}
\end{table}

\begin{figure}
\centering
\includegraphics[width=\columnwidth]{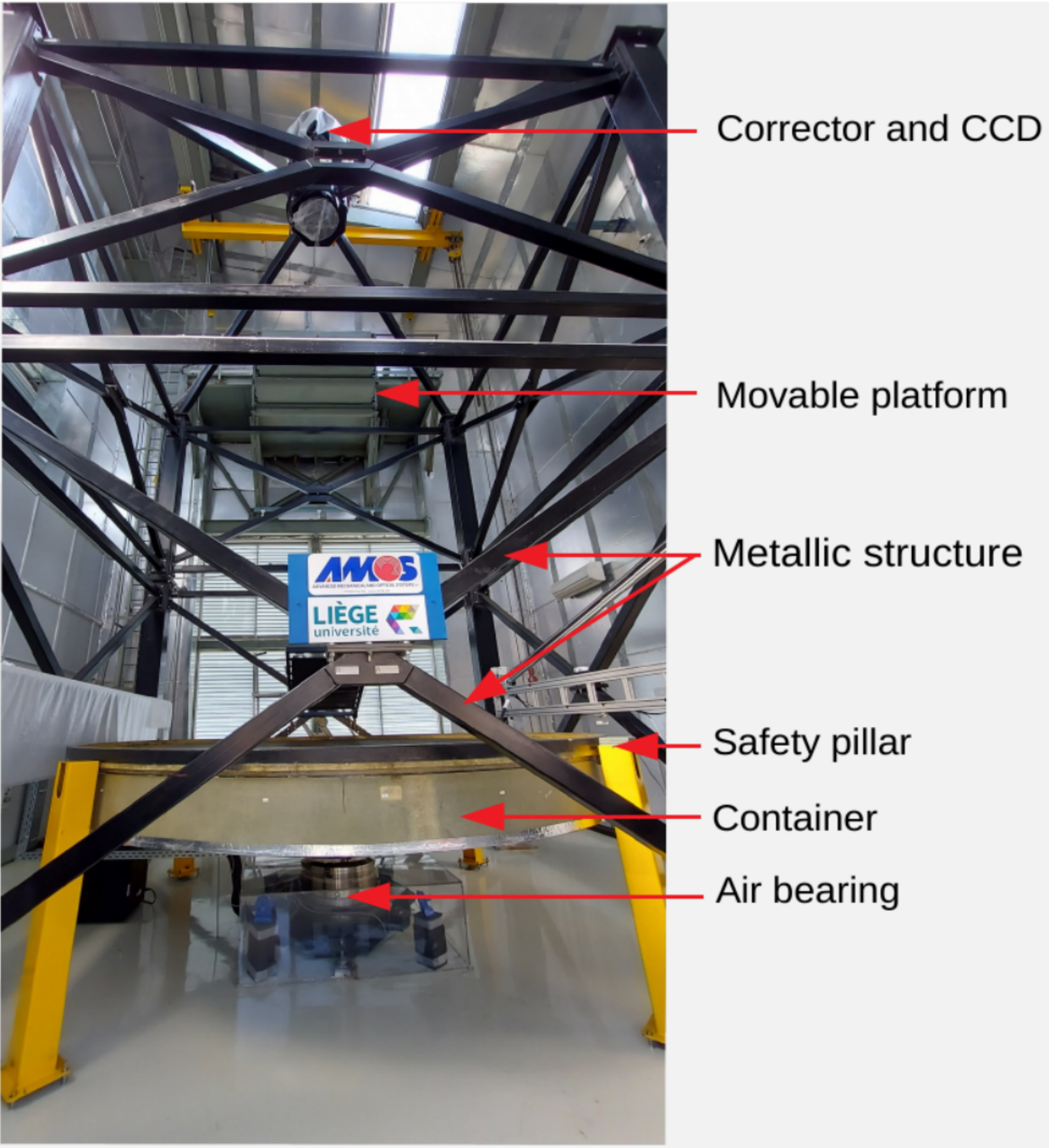}
\caption{Inside view of the fully installed ILMT. Major components are indicated. The enclosure floor is painted with high quality epoxy paint to avoid any spread outside the dome in case of accidental mercury spillage.}
\label{site-1}
\end{figure}

The ILMT building is situated in front of the 1.3m DFOT facility in Devasthal (see Fig.~\ref{site}) and it consists of three parts: compressor, data acquisition rooms and telescope enclosure. The compressor room is sufficiently away from the telescope pier to avoid passing any possible vibration to the mirror. Two air compressors and two air tanks are installed in parallel mode for an uninterrupted air supply to the air bearing. For a safe and proper functioning of the telescope, a module of pneumatic air control system, valves, air dryers, air filters and important sensors (e.g. pressure, temperature, humidity and dew-point) are also part of the system. The telescope control unit, Socabelec panel (an interface to control the filter movement and focus change), local data server and mercury monitors are installed inside the data acquisition room. The essential telescope components such as air bearing, metallic structure, optical corrector, CCD camera, mercury pumping system, etc. are disposed in the main enclosure. An inside view of the fully assembled ILMT is shown in Fig.~\ref{site-1}. Details of the ILMT installation activities are presented in \citet{Surdej-2018}.

The ILMT facility is mainly aimed at scientific projects dealing with astrometric and/or photometric variability. It will perform a deep multi-band survey of various astronomical sources in a narrow strip of sky passing over the telescope FOV. This survey will be highly useful for an independent statistical determination of the cosmological parameters (e.g. H$_{0}$, $\Omega_{M}$ and $\Omega_{\Lambda}$) by observing hundreds of multiply imaged quasars and supernovae. Furthermore, the ILMT data will be advantageous for the investigation of large scale structures, trigonometric parallaxes, photometric variability, small scale kinematics, space debris, etc. There is a long list of ILMT science cases which can be found in \citet{Claeskens-2001, Jean-2001, Surdej-2006, Surdej-2018, Finet-2013Thesis, Kumar-2014Thesis, Kumar-2015ASI, Kumar-2018MN, Kumar-2018BSRSL, Bikram-2018, Amit-2020}.

Considering the fact that mercury will be utilized to form the ILMT mirror, its safe handling is a must as mercury vapours are hazardous to the health. All possible safety precautions will be followed during the telescope operations. There will be no direct contact of the mercury to human during pumping and cleaning activities. The mercury is stored in a strong stainless steel tank and it will be transferred to the ILMT primary mirror with the help of a pumping unit. Four safety pillars are installed around the ILMT container to prevent any accidental tilt (see yellow coloured safety pillars in Fig.~\ref{site-1}). Several other equipment such as a mercury vacuum cleaner, mercury spilling kit, PPE kit, mercury vapour masks, etc. will be used for a safe handling of mercury. For a continuous monitoring of the mercury vapours, three mercury vapour detectors will be utilized.

\begin{figure*}
\centering
\includegraphics[scale=0.73]{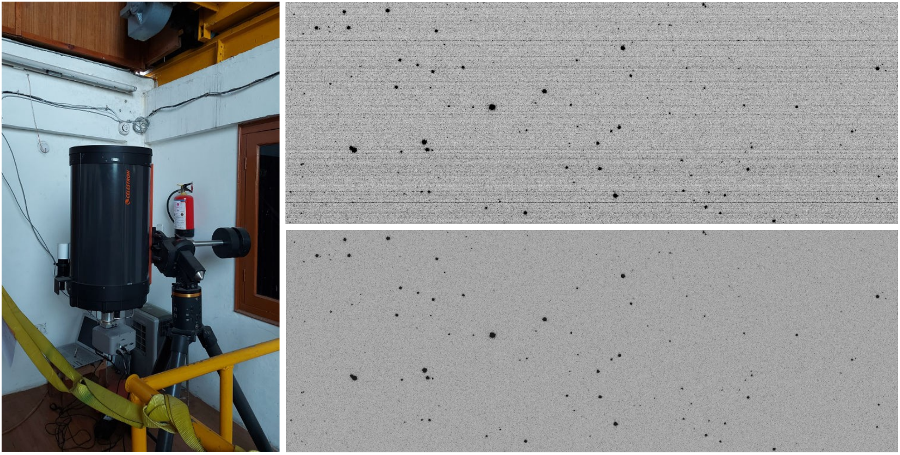}
\caption{Left panel: The 14-inch Celestron telescope installed on the 1.3m DFOT floor. TDI mode observations taken with this telescope have been used for testing the ILMT data reduction pipeline (see Section~\ref{Tel-setup}). Right panel: Cut-out images of a raw and a cleaned TDI frames observed on 23-10-2020. The upper panel shows the raw frame, whereas the lower panel shows the pre-processed frame.}
\label{14-inch}
\end{figure*}

\section{First light preparation: Data reduction pipeline testing}\label{data-test}

The installation activities of the ILMT and verification of different instruments were completed in 2019. However, its first light has been delayed due to various reasons,  the latest one being the pandemic situation. Therefore, to understand the technicalities of the data reduction, we have mounted a CCD camera on the 14-inch telescope to perform TDI mode observations. A brief description of the TDI imaging and instrument setup is provided below.  

\subsection{TDI imaging}

Zenithal telescopes such as the ILMT cannot track celestial sources like conventional telescopes. They use a special mode of observation known as the  Time-Delay Integration (TDI) technique to do the integration of photons falling from the sources \citep{Gibson-1992,Hickson-1998}. In the TDI mode, the image acquisition is performed by transferring the charges from one column to another along the right ascension direction across the CCD. The charge transfer rate is kept equal to the local sidereal rate. The two requirements for using the TDI mode technique are:

\begin{enumerate}
	\item The rate of charge transfer should match the transit rate of the sources across the detector.
	\item The source track on the CCD should be parallel to the row of the CCD.
\end{enumerate}

The TDI mode imaging has two advantages over the standard one: it provides 1) a high observing efficiency and 2) a very good flat fielding performance.
As the traditional approach in the data reduction is not sufficiently efficient and accurate for this observation, a data reduction and calibration pipeline has been developed by us to handle the data that will flow from the ILMT in the same observing mode. This pipeline can be used to reduce and analyse CCD frames obtained in the TDI mode and perform precise astrometry and aperture photometry of the detected sources. Here, we discuss the major steps involved in the reduction algorithm and some results obtained while testing the pipeline.

\subsection{TDI observations with the 14-inch telescope}\label{Tel-setup}

The observation used and envisaged for this work was based on the use of a 14-inch telescope installed just adjacent to the location of the ILMT which offers a unique opportunity to observe the same strip of sky as would be observed by the ILMT. 
The telescope was equipped with a SBIG 2k $\times$ 2k CCD (gain = 0.7 $e^{-1}$/ADU, RON = 8 $e^{-1}$ with SDSS filters $g'$, $r'$ and $i'$) capable of observing in the TDI mode. The telescope was pointed towards the zenith and was allowed to scan the sky through the $r'$-band filter. The data obtained from these observations were used to improve and optimise various algorithms and test the efficiency of the data reduction techniques.

\begin{figure*}
\centering
\includegraphics[scale=0.67]{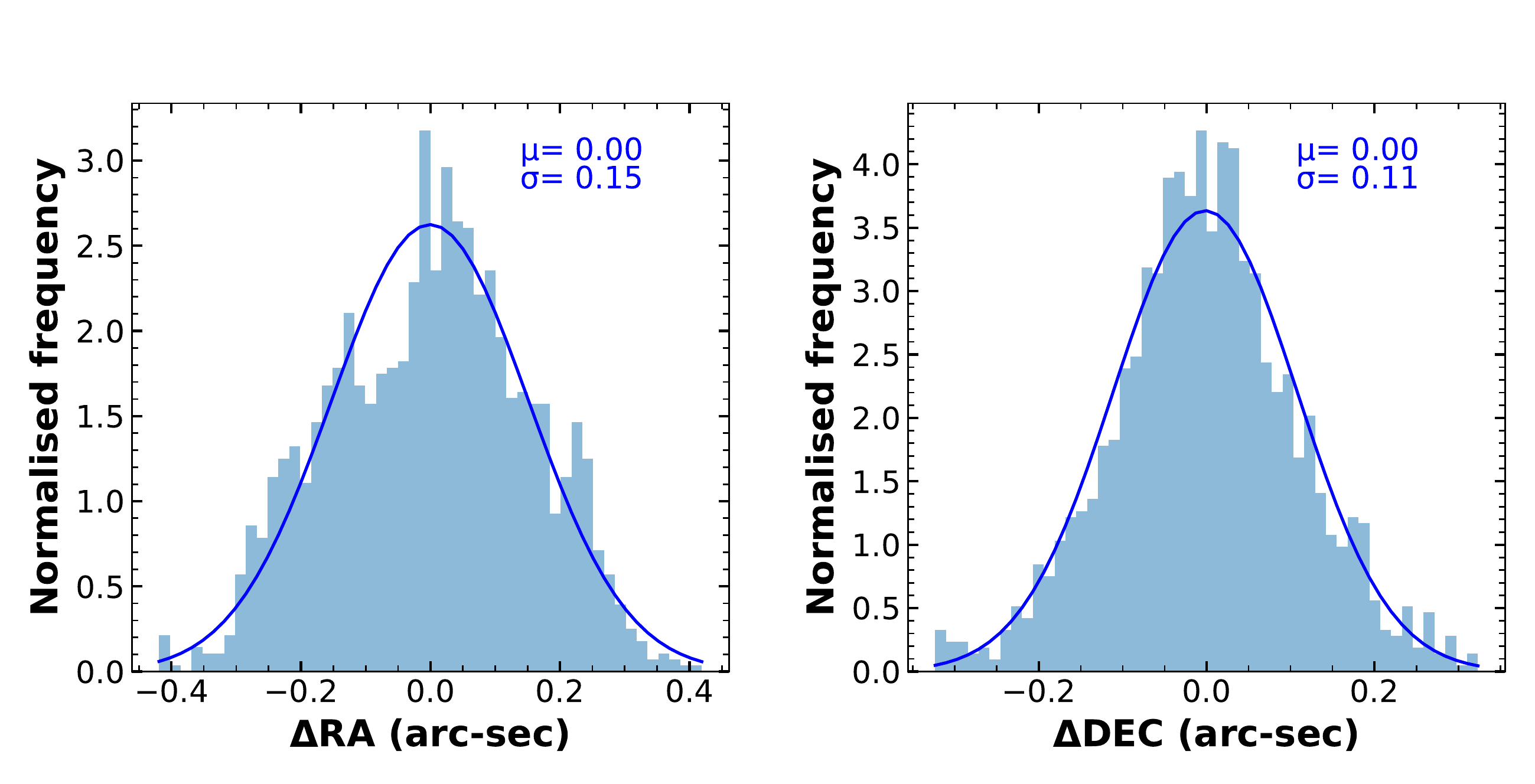}
\caption{Left panel: Normalised frequency distribution of the offsets in the calculated right ascensions of the sources compared with the GAIA coordinates. For each bin, the normalised frequency is calculated by dividing it's raw count with the total number of counts and the bin width so that the area under the histogram integrates to 1. The solid blue curve indicates the Gaussian function fitted to the distribution with the estimated mean ($\mu$) and standard deviation ($\sigma$) of the best fit given at the top right corner. Right panel: Same as the left panel but for the declinations.} \label{fig:Figure1}
\end{figure*}

\section{Data pipeline methodology}

The designed data reduction algorithm involves two steps, (i) pre-processing of the CCD data to take care of the instrumental effects and (ii) the post-processing to facilitate the scientific analysis of the data. The pre-processing includes dark subtraction, flat-field correction and proper sky subtraction and the post-processing includes astrometric and photometric calibration and further analysis based on different science goals. The unique feature of TDI mode observations is that the effect of dark, sensitivity (i.e. flat), and sky comes as an averaged resultant effect over the entire row instead of being pixel to pixel dependent (as in conventional steerable non-TDI mode imaging).

\subsection{Image pre-processing}

\subsubsection{Dark subtraction:}
The thermal excitation of electrons is a major source of noise in the CCD and these electrons are generated even in the absence of light. To correct for this dark effect, multiple dark frames were taken in the TDI mode each night with the CCD shutter closed and a super dark frame was created by median combining them. This super dark frame was then subtracted from the science image to get the dark subtracted image.

\begin{figure*}
\centering
\includegraphics[scale=0.67]{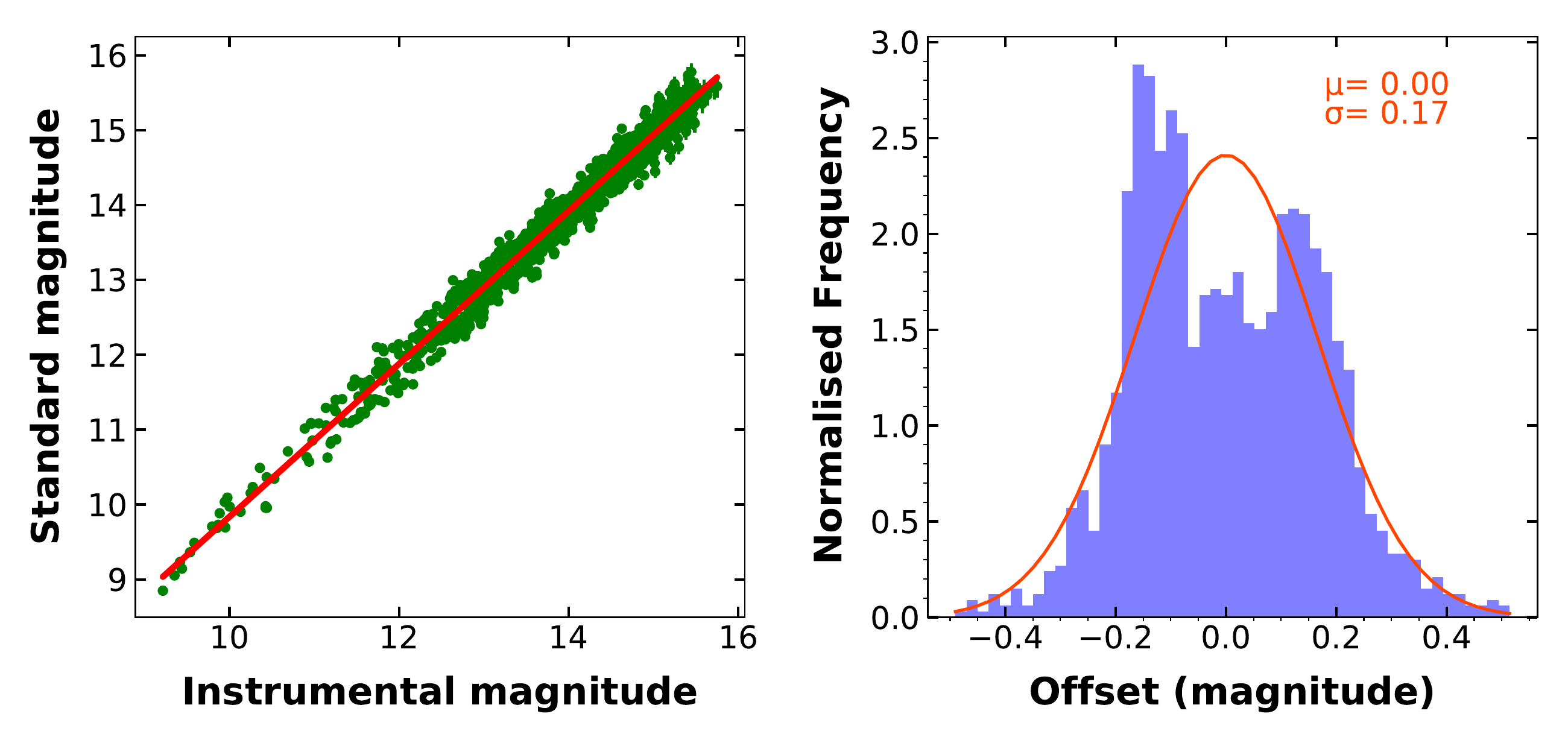}
\caption{Left panel: Standardisation of the instrumental magnitude of the sources with respect to their GAIA magnitudes. The scatter plot between the instrumental and GAIA magnitudes of the sources. The red line shows the best fit to the instrumental and standard magnitudes.  Right panel: Normalised frequency distribution of the offsets in the instrumental magnitudes with respect to the GAIA magnitudes, calculated by dividing each bin’s raw count with the total number of counts and the bin width. The orange curve shows the Gaussian function fitted to the distribution with the estimated mean ($\mu$) and standard deviation ($\sigma$) of the best fit given at the top right corner.}
\label{fig:Figure2}
\end{figure*}

\subsubsection{Flat correction:}

Flat fielding corresponds to correcting the combined CCD throughput of each pixel so that the latter responds equally to a source with the same photon flux. Flat fielding removes the effect of the pixel to pixel sensitivity variations across the array. In the dark subtracted science images, the average of the pixel values along the right ascension axis with sigma-clipping is taken in order to only include those pixels exposed to the sky so that we avoid all pixels illuminated by stars, cosmic rays and exclude the effects of bad pixels. As a result, we get 2048 values (2048 being the total number of CCD rows) which can be treated as a 1D flat. Then a lower order polynomial was fitted, which is used to divide the above 1D flat so as to get a normalised 1D flat. This normalised 1D flat (size equal to the number of rows) is used to divide each column of the dark subtracted science frame, to get the flat corrected science frame.

\subsubsection{Sky subtraction:}\label{sec:sky}

Given the large field of view (FOV) of the ILMT, it is important to do proper sky subtraction in the science image, as it can vary throughout the night. This variation could be due to the presence of the moon, clouds, dusk light, and/or twilight and can be both either temporal or spatial, or a combination of both. For this purpose, we have used our dark and flat corrected science frames and performed second order 1-dimensional polynomial fitting to the median values along the right ascension (RA) and declination (Dec) directions individually after smoothing to remove the effects of bright stars. The best fits were then subtracted from the raw frame to correct for the sky in the respective directions, which gives us a 2D sky subtracted image ready to be used for astronomical purpose.

\subsection{Image post-processing}

\subsubsection{Astrometry:}\label{sec:astrometry}
The science goals of any survey telescope rely heavily on the astrometry of the celestial sources detected in it. Knowledge of the precise celestial positions of the detected sources is required to perform the photometric calibration of the sources and further analyse them. Our astrometric measurements for the ILMT data mainly consist of four steps:
\begin{enumerate}
	\item Identification of a few targets in the fields using the plate solving engine `astrometry.net' \citep{Lang-2010}.
	\item Finding the GAIA counterparts of these sources to make use of the accurate astrometry provided by GAIA.
	\item Converting the J2000 coordinates of these sources to the observation epoch to correct for the earth precession.
	\item Calibrating the whole frame using the transformation equations given as:
\begin{equation}
    \alpha-\alpha_{0} = \frac{(y \times f_{1}-y_{0} + x \times f_{2})}{f_{3}}
    \label{eq:transformation_equations1}
\end{equation}
\vspace{-0.3in}
\begin{equation}
    \delta-\delta_{0} = \frac{(x \times g_{1}-x_{0} + y \times g_{2})}{g_{3}}
    \label{eq:transformation_equations2}
\end{equation}

where $\alpha, \delta$ are the source coordinates and x, y are the corresponding pixel positions in the CCD and $\alpha_{0}, \delta_{0}, x_{0}$ and $y_{0}$ are the same for the central pixel and $f_{1}, f_{2}, f_{3}, g_{1}, g_{2}, g_{3}$ are free parameters.
\end{enumerate}

The above mentioned procedure allows us to get a sub-arcsec astrometric accuracy in the measurement of the celestial positions of the sources detected as shown in Figure~\ref{fig:Figure1}. It displays the distribution of the offsets in the astrometric positions of the sources detected in a TDI frame using the 14-inch telescope. As can be seen, the standard deviation in the astrometry of the objects with respect to the GAIA coordinates is $\sim$\,$0.15''$ in RA and $\sim$\,$0.11''$ in DEC with mean values $\sim$\,$0''$ in both cases.

\subsubsection{Photometry and standardisation of the instrumental magnitude:}
We have used the circular aperture photometry technique with a 3 pixel radius at the locations of the objects to derive the magnitude of the detected celestial objects which corresponds to a typical seeing of around 1.3 arcsec. The aperture size can be varied depending on the seeing of the particular night and an elliptical aperture can also be selected for the extended sources such as galaxies.
Aperture photometry is the measurement of light which falls inside a well defined circular aperture. The goal is to pick an aperture which includes most of the light from the star within the aperture (i.e., after removing the background sky as discussed in Section~\ref{sec:sky}). Then the instrumental magnitude and its uncertainty were calculated, based on the measured counts and their Poisson error within the aperture.

Using the instrumental magnitude based on aperture photometry and astrometry of the image (e.g., see Section~\ref{sec:astrometry}), we searched for the apparent magnitude of the detected sources in GAIA. We naturally expect a linear relation between the instrumental and apparent magnitudes, at least for the non-varying sources. We make use of this fact to determine the zero-point and slope needed for the calibration. Plot of the instrumental and apparent magnitudes for the detected sources is shown in Figure~\ref{fig:Figure2} (after applying 3$\sigma$ clipping on the residuals i.e., instrumental magnitude -- standard magnitude), along with the best fit least squares linear fit and the distribution of the offsets detected in the calculated magnitudes with respect to the GAIA ones.

\section{Summary}

Previous LMTs were mainly used for technological demonstration and/or limited sky observations (cf. first-generation instruments and poor astronomical sites). On the other hand, the ILMT is installed at a proven high-quality astronomical site (Devasthal observatory, near Nainital, Uttarakhand). Further, it will use modern instruments to secure deep imaging of the zenith sky. Various experiments have been performed to overcome  possible failures during regular observations. The safety considerations to handle the mercury have also been pointed out. The facility is mainly aimed at science cases related to photometric and astrometric variability through the survey of a narrow sky strip ($\sim$ half a degree) passing over the ILMT. It will also provide interesting targets for the other larger observing facilities.

A data reduction pipeline has been developed to analyse the ILMT data in  real-time. To test this pipeline, we have used TDI mode observational data obtained with a 14-inch telescope installed near the ILMT location. Small tweaking was implemented while running the pipeline on the individual image frames. Its robustness still needs to be further examined. We are very optimistic to see first light of the ILMT during early 2022.

\section*{Acknowledgements}
We thank the members of the ILMT team who provided their sincere support for this project which will see its first light soon. The authors also thank the referee for insightful suggestions.

\vspace{-1em}
\bibliography{ILMT}

\begin{thebibliography}{}
\expandafter\ifx\csname natexlab\endcsname\relax\def\natexlab#1{#1}\fi

\bibitem[{{Borra}(1982)}]{Borra-1982}
{Borra}, E.~F. 1982, \jrasc, 76, 245

\bibitem[{{Borra}(1995)}]{Borra-1995}
---. 1995, \cajph, 73, 109

\bibitem[{{Borra} {$et~al$.}(1989){Borra}, {Content}, {Drinkwater}, \&
  {Szapiel}}]{Borra-1989}
{Borra}, E.~F., {Content}, R., {Drinkwater}, M.~J., \& {Szapiel}, S. 1989,
  \apjl, 346, L41

\bibitem[{{Borra} {$et~al$.}(1992){Borra}, {Content}, {Girard}, {Szapiel},
  {Tremblay}, \& {Boily}}]{Borra-1992}
{Borra}, E.~F., {Content}, R., {Girard}, L., {$et~al$.} 1992, \apj, 393, 829

\bibitem[{{Claeskens} {$et~al$.}(2001){Claeskens}, {Jean}, \&
  {Surdej}}]{Claeskens-2001}
{Claeskens}, J.~C., {Jean}, C., \& {Surdej}, J. 2001, in Astronomical Society
  of the Pacific Conference Series, Vol. 239, Microlensing 2000: A New Era of
  Microlensing Astrophysics, ed. J.~W. {Menzies} \& P.~D. {Sackett}, 373

\bibitem[{{Finet}(2013)}]{Finet-2013Thesis}
{Finet}, F. 2013, PhD thesis, University of Li{\`e}ge, Belgium

\bibitem[{{Gibson}(1991)}]{Gibson-1991-LM}
{Gibson}, B.~K. 1991, \jrasc, 85, 158

\bibitem[{{Gibson} \& {Hickson}(1991)}]{Gibson-1991}
{Gibson}, B.~K., \& {Hickson}, P. 1991, in Astronomical Society of the Pacific
  Conference Series, Vol.~21, The Space Distribution of Quasars, ed.
  D.~{Crampton}, 80

\bibitem[{{Gibson} \& {Hickson}(1992)}]{Gibson-1992}
{Gibson}, B.~K., \& {Hickson}, P. 1992, \mnras, 258, 543

\bibitem[{{Hickson} {$et~al$.}(1994{\natexlab{a}}){Hickson}, {Borra},
  {Cabanac}, {Content}, {Gibson}, \& {Walker}}]{Hickson-1994ApJ}
{Hickson}, P., {Borra}, E.~F., {Cabanac}, R., {$et~al$.} 1994{\natexlab{a}},
  \apjl, 436, L201

\bibitem[{{Hickson} \& {Racine}(2007)}]{Hickson-2007PASP}
{Hickson}, P., \& {Racine}, R. 2007, \pasp, 119, 456

\bibitem[{{Hickson} \& {Richardson}(1998)}]{Hickson-1998}
{Hickson}, P., \& {Richardson}, E.~H. 1998, \pasp, 110, 1081

\bibitem[{{Hickson} {$et~al$.}(1994{\natexlab{b}}){Hickson}, {Walker}, {Borra},
  \& {Cabanac}}]{Hickson-1994}
{Hickson}, P., {Walker}, G.~A., {Borra}, E.~F., \& {Cabanac}, R.
  1994{\natexlab{b}}, in Society of Photo-Optical Instrumentation Engineers
  (SPIE) Conference Series, Vol. 2199, Advanced Technology Optical Telescopes
  V, ed. L.~M. {Stepp}, 922--927

\bibitem[{{Hickson} {$et~al$.}(2007){Hickson}, {Pfrommer}, {Cabanac}, {Crotts},
  {Johnson}, {de Lapparent}, {Lanzetta}, {Gromoll}, {Mulrooney}, {Sivanandam},
  \& {Truax}}]{Hickson-2007LZT}
{Hickson}, P., {Pfrommer}, T., {Cabanac}, R., {$et~al$.} 2007, \pasp, 119, 444

\bibitem[{{Jean} {$et~al$.}(2001){Jean}, {Claeskens}, \& {Surdej}}]{Jean-2001}
{Jean}, C., {Claeskens}, J.~F., \& {Surdej}, J. 2001, in Astronomical Society
  of the Pacific Conference Series, Vol. 237, Gravitational Lensing: Recent
  Progress and Future Go, ed. T.~G. {Brainerd} \& C.~S. {Kochanek}, 423

\bibitem[{{Kumar}(2014)}]{Kumar-2014Thesis}
{Kumar}, B. 2014, PhD thesis, University of Li{\`e}ge, Belgium

\bibitem[{{Kumar} {$et~al$.}(2018{\natexlab{a}}){Kumar}, {Pandey}, {Pandey},
  {Hickson}, {Borra}, {Anupama}, \& {Surdej}}]{Kumar-2018MN}
{Kumar}, B., {Pandey}, K.~L., {Pandey}, S.~B., {$et~al$.} 2018{\natexlab{a}},
  \mnras, 476, 2075

\bibitem[{{Kumar} {$et~al$.}(2018{\natexlab{b}}){Kumar}, {Pandey}, {Pandey},
  {Anapuma}, \& {Surdej}}]{Kumar-2018BSRSL}
{Kumar}, B., {Pandey}, S.~B., {Pandey}, K.~L., {Anapuma}, G.~C., \& {Surdej},
  J. 2018{\natexlab{b}}, Bulletin de la Societe Royale des Sciences de
  Li{\`e}ge, 87, 80

\bibitem[{{Kumar} {$et~al$.}(2015){Kumar}, {Surdej}, {Hickson}, {Borra},
  {Finet}, {Swings}, {Habraken}, \& {Pandey}}]{Kumar-2015ASI}
{Kumar}, B., {Surdej}, J., {Hickson}, P., {$et~al$.} 2015, in Astronomical
  Society of India Conference Series, Vol.~12, Astronomical Society of India
  Conference Series, 149--150

\bibitem[{{Kumar} {$et~al$.}(2018{\natexlab{c}}){Kumar}, {Omar}, {Maheswar},
  {Pandey}, {Sagar}, {Uddin}, {Sanwal}, {Bangia}, {Kumar}, {Yadav}, {Sahu},
  {Pant}, {Reddy}, {Gupta}, {Chand}, {Pandey}, {Joshi}, {Jaiswar}, {Nanjappa},
  {Purushottam}, {Yadav}, {Sharma}, {Pandey}, {Joshi}, {Joshi}, {Lata},
  {Mehdi}, {Misra}, \& {Singh}}]{Brij-2018}
{Kumar}, B., {Omar}, A., {Maheswar}, G., {$et~al$.} 2018{\natexlab{c}},
  Bulletin de la Societe Royale des Sciences de Li{\`e}ge, 87, 29

\bibitem[{{Lang} {$et~al$.}(2010){Lang}, {Hogg}, {Mierle}, {Blanton}, \&
  {Roweis}}]{Lang-2010}
{Lang}, D., {Hogg}, D.~W., {Mierle}, K., {Blanton}, M., \& {Roweis}, S. 2010,
  \aj, 139, 1782

\bibitem[{{Mailly}(1872)}]{Mailly-1872}
{Mailly}, E. 1872, {De l'astronomie dans l'Academie royale de Belgique. Rapport
  seculaire (1772-1872)}, 99

\bibitem[{{Mandal} {$et~al$.}(2020){Mandal}, {Pradhan}, {Surdej}, {Stalin},
  {Sagar}, \& {Mathew}}]{Amit-2020}
{Mandal}, A.~K., {Pradhan}, B., {Surdej}, J., {$et~al$.} 2020, Journal of
  Astrophysics and Astronomy, 41, 22

\bibitem[{{Pfrommer} \& {Hickson}(2012)}]{Pfrommer-2012}
{Pfrommer}, T., \& {Hickson}, P. 2012, in Society of Photo-Optical
  Instrumentation Engineers (SPIE) Conference Series, Vol. 8447, Adaptive
  Optics Systems III, ed. B.~L. {Ellerbroek}, E.~{Marchetti}, \& J.-P.
  {V{\'e}ran}, 844719

\bibitem[{{Pfrommer} \& {Hickson}(2014)}]{Pfrommer-2014}
{Pfrommer}, T., \& {Hickson}, P. 2014, \aap, 565, A102

\bibitem[{{Pfrommer} {$et~al$.}(2009){Pfrommer}, {Hickson}, \&
  {She}}]{Pfrommer-2009}
{Pfrommer}, T., {Hickson}, P., \& {She}, C.-Y. 2009, \grl, 36, L15831

\bibitem[{{Pradhan} {$et~al$.}(2018){Pradhan}, {Delchambre}, {Hickson},
  {Akhunov}, {Bartczak}, {Kumar}, \& {Surdej}}]{Bikram-2018}
{Pradhan}, B., {Delchambre}, L., {Hickson}, P., {$et~al$.} 2018, Bulletin de la
  Societe Royale des Sciences de Liege, 87, 88

\bibitem[{{Sagar} {$et~al$.}(2019){Sagar}, {Kumar}, \& {Omar}}]{Sagar-2019}
{Sagar}, R., {Kumar}, B., \& {Omar}, A. 2019, Current Science, 117, 365

\bibitem[{{Sagar} {$et~al$.}(2012){Sagar}, {Kumar}, {Omar}, \&
  {Pandey}}]{Sagar-2012}
{Sagar}, R., {Kumar}, B., {Omar}, A., \& {Pandey}, A.~K. 2012, in Society of
  Photo-Optical Instrumentation Engineers (SPIE) Conference Series, Vol. 8444,
  Ground-based and Airborne Telescopes IV, ed. L.~M. {Stepp}, R.~{Gilmozzi}, \&
  H.~J. {Hall}, 84441T

\bibitem[{{Surdej} {$et~al$.}(2006){Surdej}, {Absil}, {Bartczak}, {Borra},
  {Chisogne}, {Claeskens}, {Collin}, {De Becker}, {Defr{\`e}re}, {Denis},
  {Flebus}, {Garcet}, {Gloesener}, {Jean}, {Lampens}, {Libbrecht}, {Magette},
  {Manfroid}, {Mawet}, {Nakos}, {Ninane}, {Poels}, {Pospieszalska}, {Riaud},
  {Sprimont}, \& {Swings}}]{Surdej-2006}
{Surdej}, J., {Absil}, O., {Bartczak}, P., {$et~al$.} 2006, in Society of
  Photo-Optical Instrumentation Engineers (SPIE) Conference Series, Vol. 6267,
  Society of Photo-Optical Instrumentation Engineers (SPIE) Conference Series,
  ed. L.~M. {Stepp}, 626704

\bibitem[{{Surdej} {$et~al$.}(2018){Surdej}, {Hickson}, {Borra}, {Swings},
  {Habraken}, {Akhunov}, {Bartczak}, {Chand}, {De Becker}, {Delchambre},
  {Finet}, {Kumar}, {Pandey}, {Pospieszalska}, {Pradhan}, {Sagar}, {Wertz}, {De
  Cat}, {Denis}, {de Ville}, {Jaiswar}, {Lampens}, {Nanjappa}, \&
  {Tortolani}}]{Surdej-2018}
{Surdej}, J., {Hickson}, P., {Borra}, H., {$et~al$.} 2018, Bulletin de la
  Societe Royale des Sciences de Li{\`e}ge, 87, 68

\bibitem[{{Wood}(1909)}]{Wood-1909}
{Wood}, R.~W. 1909, \apj, 29, 164

\end{thebibliography}
\end{document}